\documentclass[11pt]{article}

\usepackage{amsthm}
\usepackage{amsmath}
\usepackage{amssymb}
\usepackage{amsfonts}
\usepackage{epsfig}
\usepackage{graphicx}
\usepackage{color}
\usepackage{float}
\usepackage[caption = false]{subfig}
\usepackage{algorithm,algorithmic}
\usepackage{multirow}
\usepackage{comment}
\usepackage{hyperref}

\newtheorem{theorem}{Theorem}

\newtheorem{lemma}[theorem]{Lemma}

\newtheorem{assumption}[theorem]{Assumption}
\allowdisplaybreaks

\begin{document}
\title{\bf Estimating Incremental Acquisition of Content Launches in a Subscription Service}
\author{Hamidreza Badri \\  \href{mailto:hbadri@netflix.com}{hbadri@netflix.com} \and Alex Kaufman \\  \href{mailto:akaufman@netflix.com}{akaufman@netflix.com} }

\date{\today} 

\normalsize{}
\maketitle

\begin{abstract}
Subscription services face a difficult problem when estimating the causal impact of content launches on acquisition. Customers buy subscriptions, not individual pieces of content, and once subscribed they may consume many pieces of content in addition to the one(s) that drew them to the service. In this paper, we propose a scalable methodology to estimate the incremental acquisition impact of content launches in a subscription business model when randomized experimentation is not feasible.  Our approach uses simple assumptions to transform the problem into an equivalent question: what is the expected consumption rate for new subscribers who did \emph{not} join due to the content launch? We estimate this counterfactual rate using the consumption rate of new subscribers who joined just prior to launch, while making adjustments for variation related to subscriber attributes, the in-product experience, and seasonality. We then compare our counterfactual consumption to the actual rate in order to back out an acquisition estimate. Our methodology provides top-line impact estimates at the content / day / region grain. Additionally, to enable subscriber-level attribution, we present an algorithm that assigns specific individual accounts to add up to the top-line estimate. Subscriber-level attribution is derived by solving an optimization problem to minimize the number of subscribers attributed to more than one piece of content, while maximizing the average propensity to be incremental for subscribers attributed to each piece of content. Finally, in the absence of definitive ground truth, we present several validation methods which can be used to assess the plausibility of impact estimates generated by these methods.

 
\end{abstract}
 
\section{Introduction}

In a subscription business model, a customer pays a recurring fee, typically monthly or yearly, for unlimited access to content offered by the service. Examples include subscription video on demand companies such as Netflix and Disney+, music streaming services like Apple Music or Spotify, e-books or magazine subscription services such as Kindle Unlimited, and subscription-based fitness businesses such as Peloton. In the context of these examples, a piece of content can be considered as a new film, TV show, song, album, e-book, audiobook, or class.

Flow of new content on the service is an important growth lever for subscription services. For example, video-on-demand companies release new movies and shows every week. Content brings value to a service in two fundamental ways: they may cause current members to retain on the service more often, or they may cause new members to join the service. In this paper, we present modeling innovation designed to improve estimation of the latter impact -- the causal effect of content launches on new member acquisition. Such estimation can help inform content prioritization in programming and marketing, particularly in lower-penetrated markets where non-member content preferences may not be well understood.

Estimating the acquisition impact of content launches in subscription services is challenging, mainly due to three factors. First, most existing attribution models are focused on the problem of marketing attribution. A customer purchases a product and the firm wishes to attribute that purchase to some combination of previous events, such as ad impressions.  The firm uses cookies or other forms of tracking to observe the timing and nature of previous events to which the customer was exposed.  The firm also observes previous events for potential customers who did not purchase. Then the attribution is done via training a model on data with binary “convert” / “don’t convert” target variables. However, in the context of a service where subscribers must join in order to consume content, all observed outcomes are "convert." For example, marketers can observe ad exposure for both converters and non-converters, but subscription services cannot easily observe the behavior of non-subscribers, i.e. what specific content they would have consumed if they had decided to subscribe. Hence, launch attribution faces a severe data censoring problem which makes the marketing attribution literature challenging to adapt to our context. 

Second, randomized experiments tend to be impractical for determining the incremental impact of content launches, since subscription services generally launch each piece of content only once, and after being launched, it can be accessed and consumed by all subscribers. Inability to use randomized experiments makes causal identification and validation challenging.

Third, subscription services face the additional hurdle that, once a member has subscribed, all content becomes costless to consume. This means that subscribers may quickly consume a large amount of content, including many pieces of content that were unrelated to their decision to join, making it difficult to pick out which particular content, if any, was pivotal to that decision.  

This paper introduces the acquisition impact model (AIM), a model that uses observational data to estimate the number of sign ups caused by a piece of content at its launch.  AIM combines the intuition behind first touch attribution -- that members quickly consume the content that has caused them to sign up -- with a new methodology that compares cohorts of new members based on their sign up date relative to the content launch date.  Specifically, we use the pre-launch sign up cohort as a covariate-adjusted control group for the post-launch cohort, and compare their consumption behavior in order to tease out the number of incremental sign ups that exist due to the new content launch. Under the assumptions that a) all incremental sign ups quickly start consuming the content they signed up for, and b) the underlying preferences of new subscribers who are non-incremental for the newly launched content are smooth around the launch date, our method delivers plausibly-causal estimates.

The contributions of the paper are: (1) present a scalable model to estimate the number of sign ups caused by a new content at its launch (2) propose a framework to isolate specific subscribers that signed up for the new content and (3) propose several ways to validate the methodology using external sources of truth.

The paper is organized as follows. Section 2 shows how, under plausible assumptions, estimating incremental subscribers is equivalent to estimating the non-incremental consumption rate, then shows how we might adjust the stream rate to account for differences in covariates between populations. Section 3 describes a methodology that can be used to identify subscribers who signed up for the new content from those who streamed the content due to seasonlity factors, higher user activity or product promotion. Section 4 proposes several methodologies we could use to validate the estimates. Section 5 reviews the attribution literature, and Section 6 concludes.

\section{Estimation Methodology}
This section sets up AIM and shows that if we assume incremental subscribers always consume the content that causes them to subscribe upon signing up, we can solve for the number of incremental new subscribers using the non-incremental consumption rate. We then discuss covariates that may influence the probability of consumption among non-incremental subscribers, and how we can alter covariate distributions to get adjusted consumption rates.

\subsection{Setup}
Suppose we observe a population $N_t$ of new subscribers on a particular day $t$:

\begin{equation}\label{eq:setup}
N_t=N_{\text{Catalog}}+\sum_{j\in \mathbb{A}}N^{+}_{jt}.
\end{equation}
$N_t$  is the sum of all subscribers who joined the subscription service specifically because of content $j$, or because of the subscription service as a whole, which we indicate by catalog-driven sign ups. A subscriber who joined due to content $j$ is “incremental for content $j$ ” while a subscriber who joined for other reasons is “non-incremental for content $j$ ”. The sum of all incremental subscribers for $j$ on day $t$  is $N^{+}_{jt}$ , while the sum of all non-incremental subscribers is $N^{-}_{jt}$ . We wish to estimate $N^{+}_{jt}$ but observe only $N_t$ . 

Information on the consumption behavior of subscribers can help to isolate $N^{+}_{jt}$. Suppose the probability that a given subscriber $i$ signing up on day $t$  consumes content $j$  can be defined by the $f(.)$ function:

$$p_{ijt} = f(X_{ijt})$$
where $p_{ijt}$ is the probability of a binary consumption outcome calculated over an appropriate time window, and $X_{ijt}$ is a set of covariates capturing the underlying preference of subscriber $i$  toward content $j$,  how content $j$  is shown to subscriber $i$  in the product,  the age of subscription $i$  at the time content $j$  becomes available to $i$, and subscriber activity over the appropriate time window defined for the consumption of content $j$. The set of subscribers who consume content $j$  is denoted $\mathbb{S}_{jt}$, the set of incremental and non-incremental subscribers for content $j$  are denoted by  $\mathbb{S}_{jt}^{+}$ and  $\mathbb{S}_{jt}^{-}$, respectively. 

\begin{assumption}\label{assumption}
	If subscriber {i} is incremental to content {j}, then $p_{ijt}=1$.
\end{assumption}

The reasoning is that for content to be pivotal to the decision to join, the subscriber must value it and be aware of it before joining. Salient, valued content is likely to be consumed.  This assumption helps identify the data elements needed to estimate $N^{+}_{jt}$ . Define the number of consumers of content $j$ who sign up on day $t$ by $S_{jt}$, then because $p_{ijt}=1$ for incremental subscribers, we have:

$$S_{jt}=\sum_{i\in \mathbb{S}_{jt}}p_{ijt}$$
$$S_{jt}=\sum_{i\in \mathbb{S}_{jt}^{+}}p_{ijt}+\sum_{i\in \mathbb{S}_{jt}^{-}}p_{ijt}$$
$$S_{jt}= N^{+}_{jt}+\sum_{i\in N^{-}_{jt}}p_{ijt}$$
$$S_{jt}= N^{+}_{jt}+N^{-}_{jt}\times \bar{p}_{ijt}$$
$$S_{jt}= N^{+}_{jt}+\left(N_t-N^{+}_{jt}\right)\times  \bar{p}_{ijt}$$

which leads to our incrementality equation:

\begin{equation}\label{eq:incrementality}
	 N^{+}_{jt} = \frac{S_{jt}-N_t\times   \bar{p}_{ijt}}{1-  \bar{p}_{ijt}}.
\end{equation}

In other words, given Assumption 1 we can recover $ N^{+}_{jt}$  from observable data $N_t$ and $S_{jt}$  by estimating the average consumption probability for non-incremental subscribers $N^{-}_{jt}$, i.e. $ \bar{p}_{ijt}$.  Because we don’t know which subscribers are non-incremental we can’t directly observe $\bar{p}_{ijt}$, which means we cannot directly solve for $N^{+}_{jt}$. However, we can make inferences about $\bar{p}_{ijt}$ using information from other populations of non-incremental subscribers.

\subsection{Control Cohort Definition}
Non-incremental subscribers are people who joined the subscription service for reasons other than content $j$, and who may or may not consume content $j$. Their chances of consumption depend on their underlying preferences toward the content, how the content is promoted to them, and so on. A plausible assumption is that, on average, non-incremental subscribers joining just before the launch of content $j$ have similar preferences toward content $j$ as non-incremental subscribers joining just after launch. Those preferences trigger in-product promotion in similar ways, and all else equal would lead to similar chances of consumption.  If this is true, we can use the consumption rate of subscribers in the pre-launch cohort as a ``control'' group to estimate $\bar{p}_{ijt}$.

The pre-launch cohort is defined as subscribers joining the service in the $T_0$ days prior to launch. The choice of $T_0$ depends on the trade-off between the benefit of extra pre-launch data versus the need to make more extreme covariate adjustments due to time lag between control and treatment cohorts. We may also wish to end the window several days before launch, in order to exclude the possibility of incremental new members that join in the immediate run-up to launch. 

We wish to estimate the non-incremental subscriber consumption probability in $\mathbb{S}_{jt}$ , but $\mathbb{S}_{jt}$ is a mixture of incremental and non-incremental subscribers. However, we also observe a different population of subscribers $\mathbb{B}_j$, e.g. pre-launch cohort, which we assume contains no incremental subscribers for content $j$ since content $j$ was not available at the time the cohort joined. 

\begin{assumption}\label{assumption}
	For subscribers joining in the $T_0$-day period prior to the launch of content $j$, $\mathbb{B}^{+}_j = 0$ and $\mathbb{B}^{-}_j = \mathbb{B}_j$ 
\end{assumption}

We can use $\mathbb{B}_j$ in several ways. First if $\mathbb{B}_j$ and $\mathbb{S}_{jt}^-$ have the same distribution of covariates, i.e. $X_{ijt}$, then we can simply use the average stream rate for subscribers in $\mathbb{B}_j$ to estimate $\bar{p}_{ijt}$:
$$\bar{p}_{ijt}=\frac{\sum_{i \in \mathbb{B}_j }p_{ijt}}{|\mathbb{B}_j|}$$
and then use \eqref{eq:incrementality} to estimate $N^{+}_{jt}$. 

However, the assumption that  $\mathbb{B}_j$ and $\mathbb{S}_{jt}^-$ have the same distribution $X_{ijt}$ of covariates is unlikely to hold in practice. When pre-launch and post-launch cohorts differ in important ways other than the presence of incremental subscribers, we can make adjustments to account for these differences. For instance, pre-launch subscriptions are necessarily older than post-launch subscriptions. If subscription age affects the probability of consumption, we can learn this relationship and use it to adjust for age differences between the cohorts. Similarly, if there are differences between cohorts in how content is presented in the product, or in subscriber activity upon sign up, we can adjust for those differences as well. 

As before, there is a relationship between $p_{ijt}$ and various covariates such as subscriber age, subscriber activity, content promotion, and underlying preference given by

$$p_{ijt} = f(X_{ijt}).$$

Unlike preference toward content $j$ -- which is not directly observable but which for non-incremental subscribers is assumed to be continuous across the content launch boundary -- age, subscriber activity, and promotion are observable to us. We can adjust for them together by estimating a model that explains the relationship between these factors and $p_{ijt}$:

\begin{eqnarray}\label{eq:ml_model}
	p_{ijt} = \hat{f}\left(\text{age}_{ijt},\text{activity}_{ijt},\text{promotion}_{ijt}\right)
\end{eqnarray}
We can train such a model using the pre-launch cohort on the pool of all available content launches. Model \eqref{eq:ml_model} provides an adjusted consumption probability for $\mathbb{B}_j$ that simulates the probability of consumption if $\mathbb{B}_j$ had the same joint distribution of age, subscriber activity, and  promotion as  $\mathbb{S}_{jt}^-$. Finally, we substitute this adjusted consumption rate into \eqref{eq:incrementality} to solve for $N^{+}_{jt}$. 

Theoretically, we could extend the post-launch window indefinitely in order to capture more long-tail incremental subscribers. The limiting factors are the comparability of new subscribers to an increasingly-distant pre-launch cohort and the fact that measurement error in the tail may overwhelm the trickle of true incremental subscribers.

\subsection{Example}
To be concrete, imagine a streaming music service that releases a new album A. We wish to estimate how many new subscribers joined the service specifically because of the launch of A. We define a subscriber ``consuming'' A as streaming more than a certain percentage of the album within a certain window of the album becoming available to that subscriber. For simplicity imagine that promotion of A is the same for all subscribers, and that the only reason different cohorts of new subscribers may differ in their consumption rate of A is the presence or absence of incremental subscribers who specifically joined the service in order to stream A.

\begin{figure}
	\centering
	\includegraphics[width=170mm]{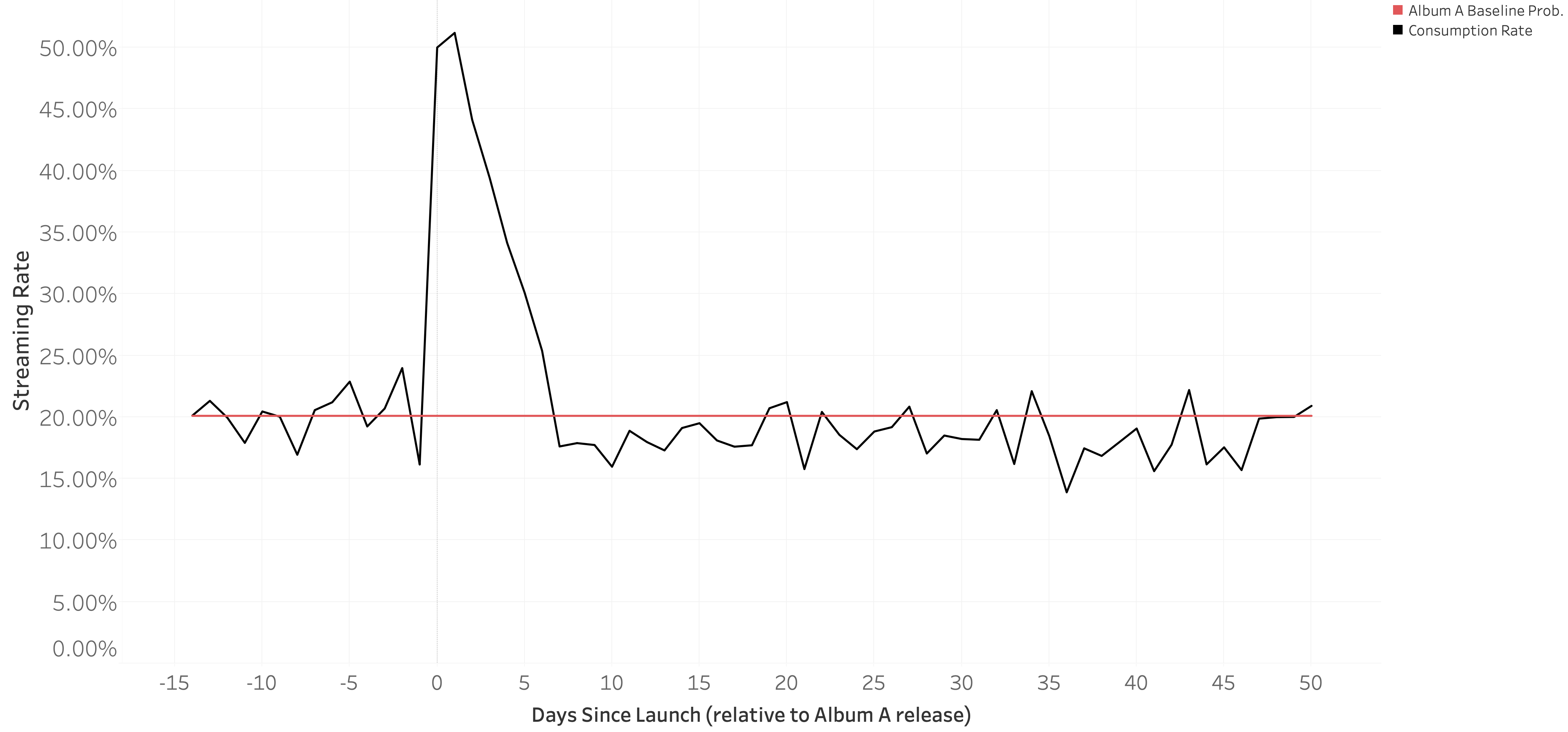}
	\caption{Simulated non-incremental streaming rate as a function of join date in a toy example}
	\label{pic:baseline_watch}
\end{figure}

For new subscribers who joined prior to the launch of A, the first opportunity to stream A was the day of A's launch. For new subscribers who joined post-launch, the first opportunity to stream A was the first day of their new subscription. Figure \ref{pic:baseline_watch} charts the consumption rate for subscribers who joined on different dates. We observe the consumption rate spike for subscribers joining post-launch. In our toy example, because covariates are assumed to be balanced for cohorts across the date of launch, this increase in consumption must derive from the presence of subscribers who joined because of A and stream A with probability 1, rather than baseline probability $p_{ijt} \approx 0.2$.

\begin{figure}
	\centering
	\includegraphics[width=170mm]{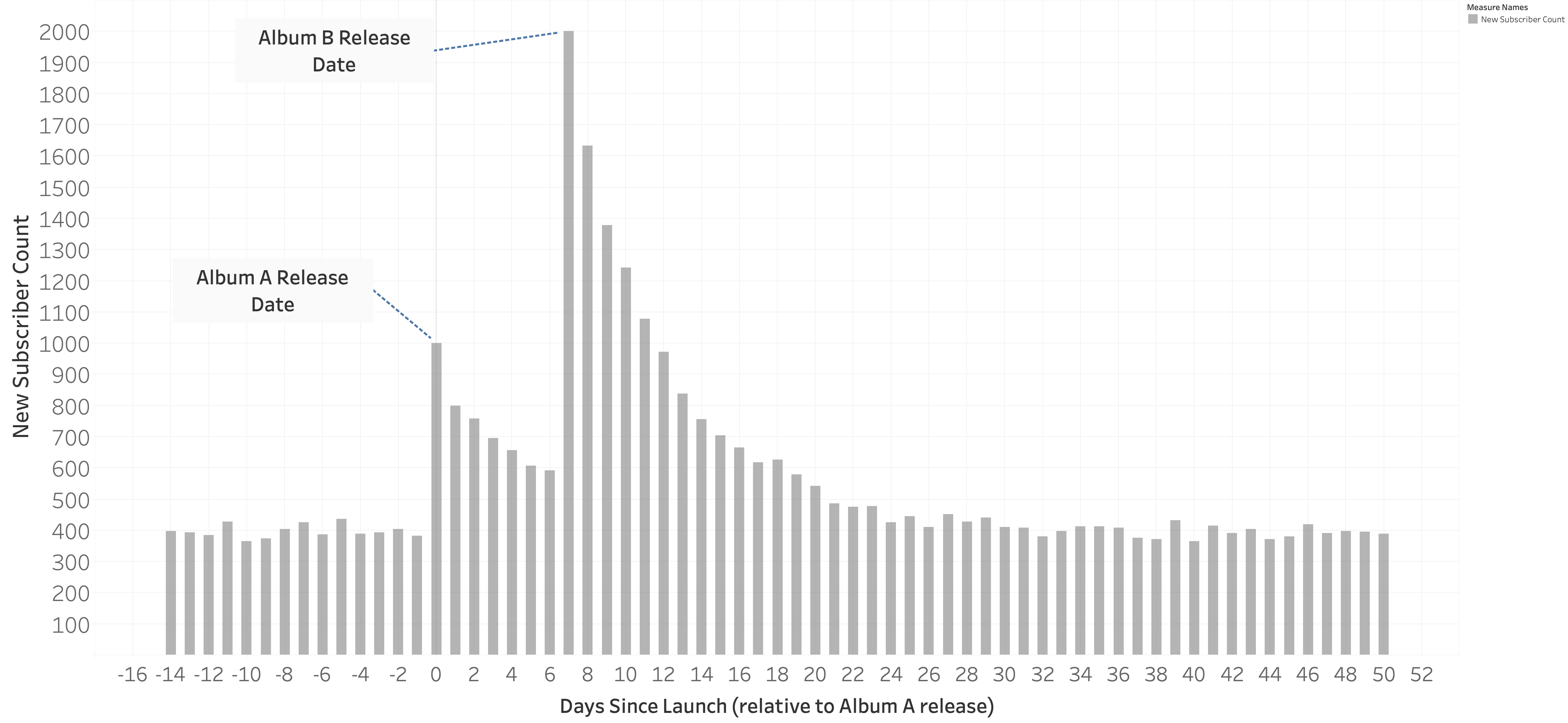}
	\caption{Simulated new subscriber count as a function of join date in a toy example}
	\label{pic:input_data}
\end{figure}

Figure \ref{pic:input_data}  plots the number of new subscribers each day. On launch day 1000 new subscribers joined the service, of whom 500 streamed the new album. Using the pre-launch cohort, we estimate that the streaming rate among non-incremental new subscribers is 20\%. 

We can then use Equation \eqref{eq:incrementality} to solve for $N^{+}_{jt}$:
$$N^{+}_{jt} = \frac{S_{jt}-N_t\times   \bar{p}_{ijt}}{1-  \bar{p}_{ijt}}$$
$$N^{+}_{jt} = \frac{500-1000\times   0.2}{1-  0.2} = 375.$$

We can perform a similar calculation every day post launch, and take the area under the curve to find the incremental acquisition impact of album A.

\begin{figure}
	\centering
	\includegraphics[width=170mm]{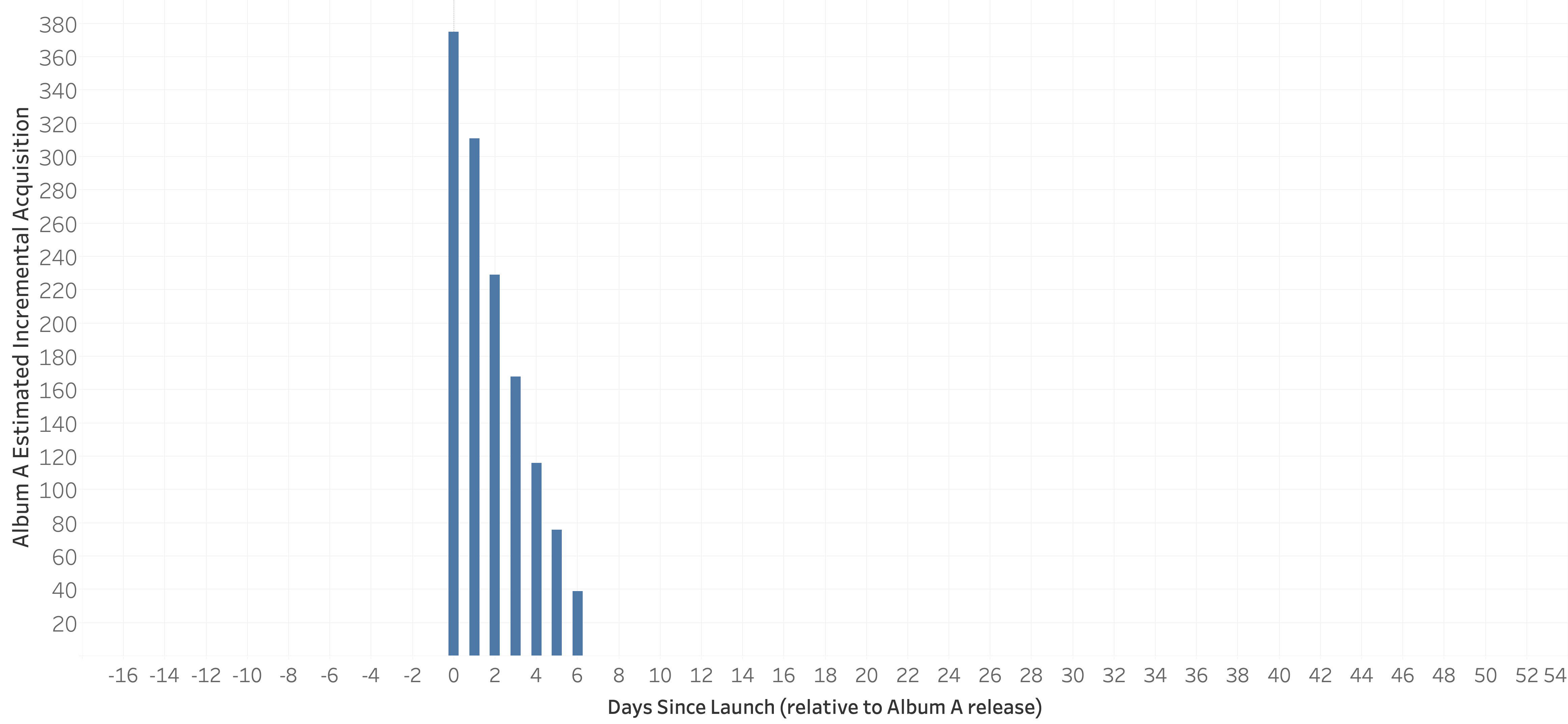}
	\caption{Estimated incremental acquisition for Album A}
	\label{pic:output_data}
\end{figure}

Note that even though there was a second rise in new subscribers a week after launch, because those new subscribers were not more likely than average to stream album A, they are not attributed to A. The second hump may be due to album B, or to a transient factor unrelated to content launches. In general, AIM can be used to estimate the impact of multiple overlapping content launches, provided that preferences for A and B are not strongly correlated.

\section{Subscriber-Level Attribution}
The previous section explained how the total number of sign ups (“acquisition impact”) caused by a content launch can be estimated by forming a synthetic control. However, this process only yields an aggregate acquisition count -- it does not tell us which particular subscribers, among those who consumed the content, were most likely to have signed up because of the launch. Subscriber-level attribution is necessary to answer any questions about the attributes, preferences, and subsequent activity of those who are drawn in by the content. For instance, such attribution is necessary to study users who signed up for specific content but later churned due to deficiency in similar content that might have retained them.

In this section, we describe a framework for subscriber-level attribution. We start by discussing an important property of the incremental subscribers.

\begin{lemma} \label{lemma_pij}
	Let the non-incremental consumption probability of subscriber $i$ and content $j$ estimated by \eqref{eq:ml_model} be $\hat{p}_{ijt}$, then we have:
	$$\frac{\partial P(\text{subscriber } i \text{ being incremental to content } j)}{\partial \hat{p}_{ijt}}\le0.$$
\end{lemma}

The proof is in Appendix \ref{lemma_pij_proof}. Lemma \ref{lemma_pij} states that for subscribers who consume content $j$, the probability of being incremental to content $j$ is decreasing in the baseline probability of consumption $\hat{p}_{ijt}$. This helps us to find a rank order to better identify the possible incremental subscribers for each piece of content, e.g. we can use $1-\hat{p}_{ijt}$ to rank subscribers according to their probability of being incremental for $j$. 

Note that we explicitly incorporate subscriber level features when estimating $\hat{p}_{ijt}$, such as subscriber activity. For example, if $\hat{p}_{ijt}$ is increasing with respect to subscriber activity, then it is more plausible to attribute low-activity subscribers as incremental subscribers of content $j$, i.e. new sign ups with higher activity (who consume many pieces of content) will have a lower likelihood of being incremental for any given piece of content that they consume. 

\begin{figure}
	\centering
	\includegraphics[width=150mm]{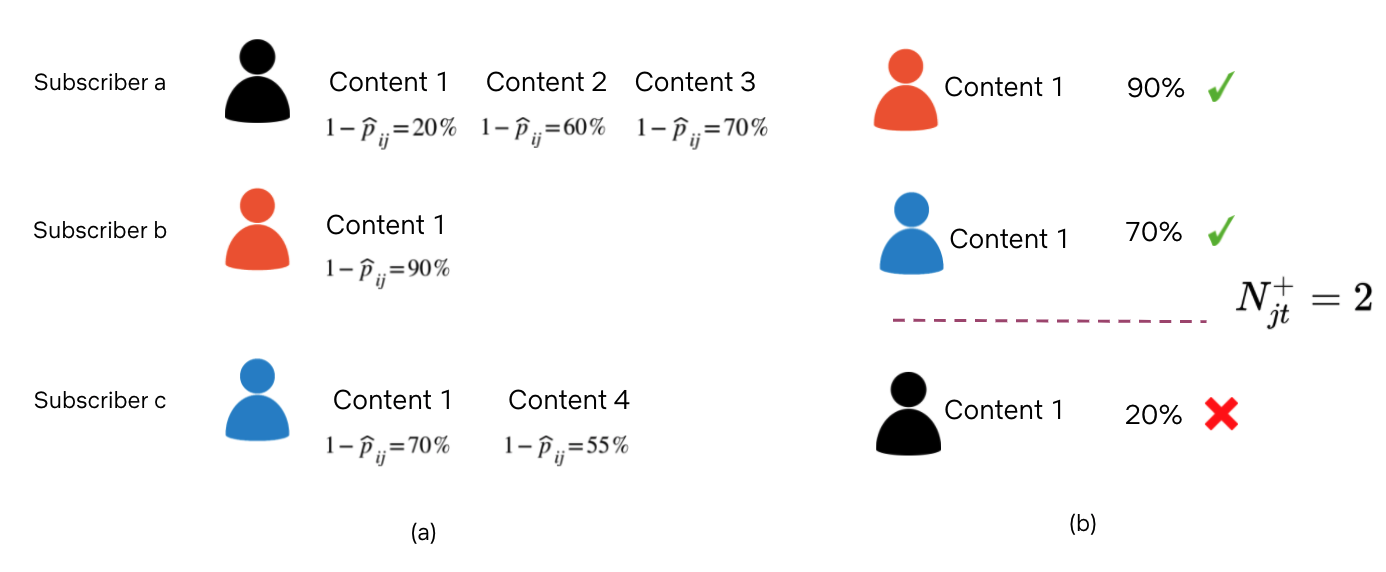}
	\caption{(a) Each subscriber consumes a different number of contents upon signing up. Numbers reported in the graph are $1-\hat{p}_{ijt}$ for a hypothetical example. (b) Ranked order incremental likelihood attribution example for Content 1.}
	\label{pic:account_attribution}
\end{figure}

After estimating the $1-\hat{p}_{ijt}$ for subscriber-content pairs (Figure \ref{pic:account_attribution}-a), we can limit the number of incremental subscribers being attributed to each piece of content by  $N^{+}_{jt}$. There are various ways which we can rank different subscribers for each piece of content. For example, if the aggregate model estimates two incremental acquisitions for content 1 on a particular day, and there are three subscribers consuming content 1 upon signing up, we can choose the two subscribers with the highest affinity as the incremental subscribers for content 1 (Figure \ref{pic:account_attribution}-b). Another possibility is to decay $1-\hat{p}_{ijt}$ proportionally to the  order that each piece of content was consumed by each subscriber or rank order of $1-\hat{p}_{ijt}$  for each subscriber.  More specifically, if a subscriber consumes more than one newly launched content upon signing up, we can reward content that was consumed sooner after the sign up event, or we can penalize contents with lower $1-\hat{p}_{ijt}$ probability.

The main advantage of the above approaches is simplicity.  However, these approaches attribute subscribers independently across pieces of content, and don’t take into account the fact that a subscriber might be already attributed to one piece of content while attributing it to another one. For this reason, subscribers may be subject to multiple attribution. 

\subsection{Assignment Minimizing Multiple Attribution} \label{multi}

To address the issue of multiple attribution, we can consider this task as an assignment problem \cite{ramshaw2012minimum} and formulate it as a mixed linear integer problem. Written mathematically, we have:

\begin{equation}\label{optimization}
	\min_{x,y}\sum_{i} y_i -\lambda \sum_{i,j} (1-\hat{p}_{ijt}) x_{ij}
\end{equation}
subject to:
\begin{equation}\label{constrain-eia}
	\sum_{i\in S_{jt}}x_{ij}=N^+_{jt},\text { } \forall j 
\end{equation}
\begin{equation}\label{constrain-M1}
	\sum_{j\in \mathbb{T}_i}x_{ij}\le 1+My_i,\text { } \forall i
\end{equation}
\begin{equation}\label{constrain-M2}
	\sum_{j\in \mathbb{T}_i}x_{ij}\ge 2-M(1-y_i),\text { } \forall i
\end{equation}
$$x_{ij}\in\{0,1\}; y_i \in \{0,1\}$$

where $y_i$ is a binary variable that will be 1 if subscriber $i$ is attributed to more than one piece of content, $x_{ij}$ is a binary variable that will be 1 if subscriber $i$ is attributed to content $j$, $M$ is an arbitrary large number, $\mathbb{T}_i$ is the set of content being consumed by subscriber $i$ upon sign up, and $\lambda$ is a positive scalar that  defines the trade-off between 1:1 allocation and the affinity of subscribers attributed to different piece of content. 

Equation \eqref{optimization} defines the objective function for the optimization problem. Our goal is to minimize the number of subscribers that are being attributed to more than one content, while maximizing affinity of the subscribers attributed to contents. When $\lambda$ is zero, we achieve the maximum 1:1 allocation across subscribers and contents, as we increase $\lambda$, the optimization model starts penalizing solutions that allocate subscribers with low affinity, i.e. $1-\hat{p}_{ijt}$, across pieces of content. Constraint \eqref{constrain-eia} makes sure that we attribute enough subscribers to each content and day, and constraints \eqref{constrain-M1} and \eqref{constrain-M2} use disjunctive programming  \cite{balas1985disjunctive} to linearize the logical constraint of:

$$
y_i=\begin{cases}1,& \sum_{j\in \mathbb{T}_i} x_{ij}\ge2 \\
		0 ,&  \sum_{j\in \mathbb{T}_i} x_{ij}\le1 \end{cases}.
$$

\begin{figure}
	
	\subfloat[]{\includegraphics[width = 80mm]{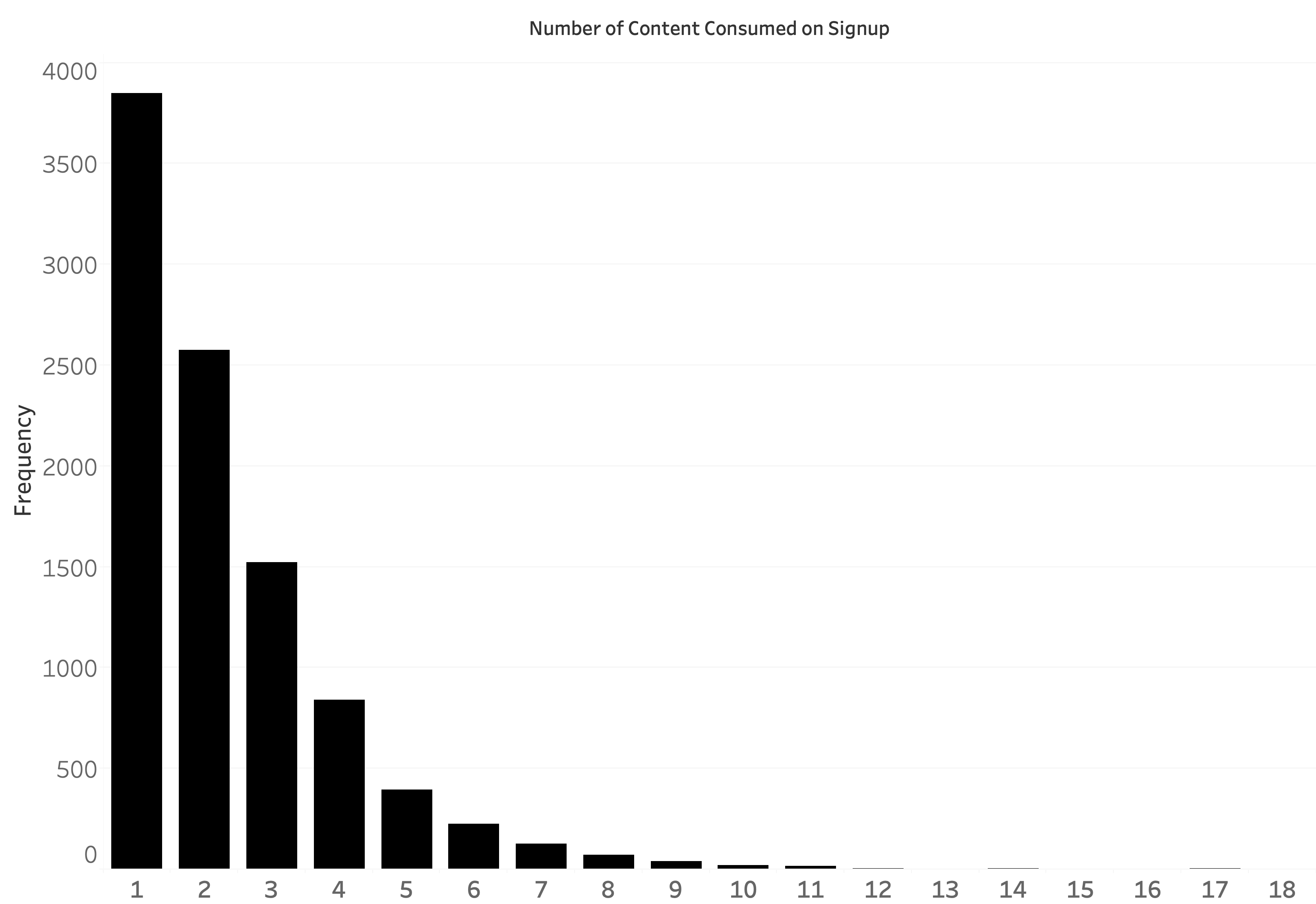}} 
	\subfloat[]{\includegraphics[width = 80mm]{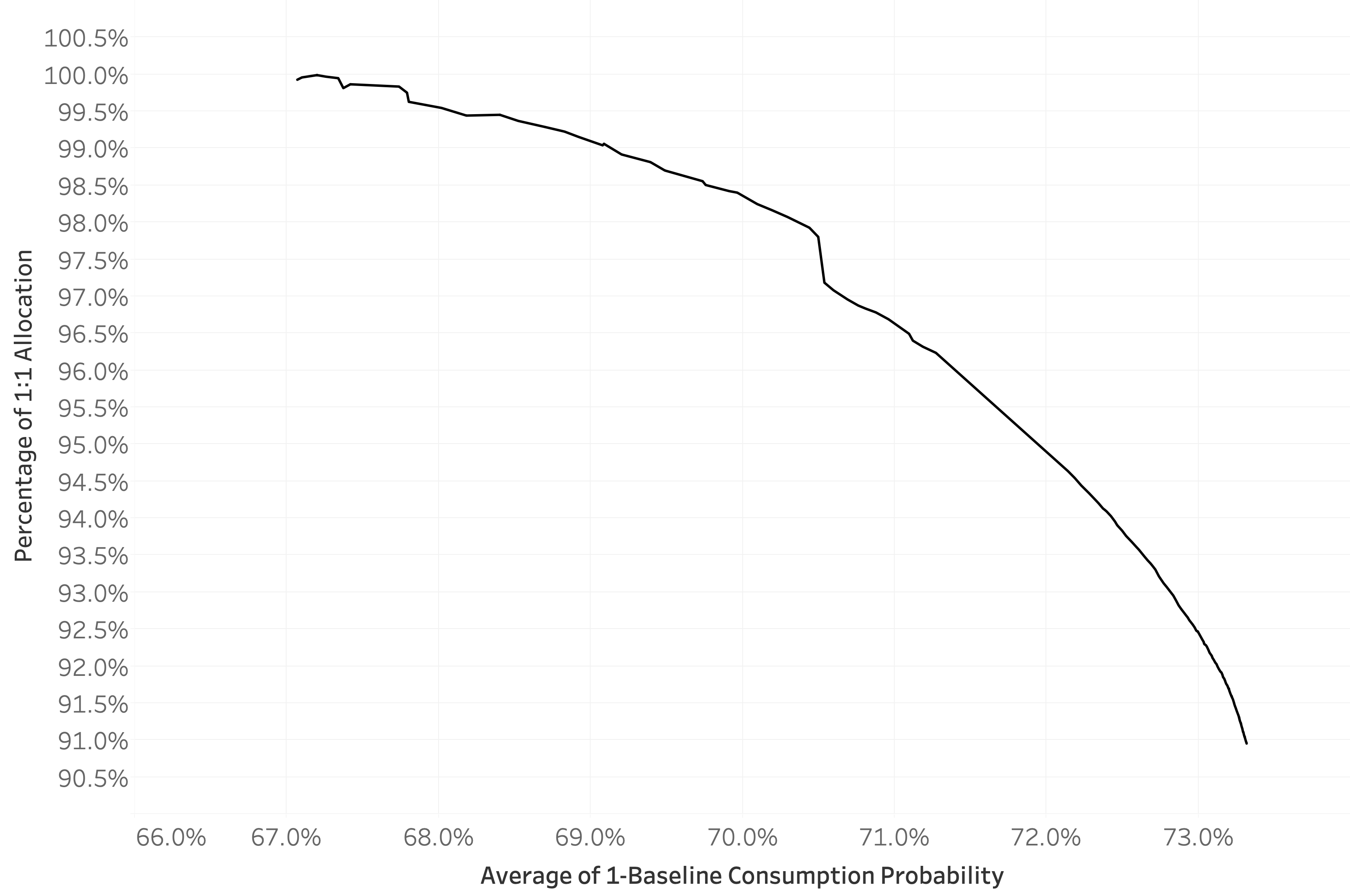}}  
	
	\caption{(a) Frequncy of number of contents being consumed by a subscriber upon signing up in the simulated example. (b) Trade-off between the reduction in percentage of 1:1 allocation and average baseline probability of subscribers attributed across contents. }
	\label{fig-pareto}
\end{figure}

Figure \ref{fig-pareto} (b) plots the trade-off between two conflicting objectives defined in \eqref{optimization}. We simulate consumption behavior of 10K new subscribers and assume that each of them has access to more than 1K pieces of contents on a subscription service. We randomly choose a set of contens that will be consumed by each subscriber on the sign up date and generate $p_{ijt}$ for each subscriber / content pair based on a uniform distribution. In this example, 60\% of subscribers consume more than one piece of content after the sign up event (Figure \ref{fig-pareto} (a)). One can observe a diminishing return in the reduction of 1:1 allocation. In particular, the allocation solution suggests a marginal improvement over 1:1 allocation for solutions with less than 70.5\% of average $1-\hat{p}_{ijt}$.

\section{Validation Methods}

The gold standard for validating model output is to compare it against casual estimates generated by a randomized experiment.  However, even when randomized experiments are infeasible we can still make progress on validation. In this section, we describe several non-experimental methods that can be used to tune the model and build confidence in its output. We recommend using an ensemble approach to validation, as over-reliance on any one approach may lead to overfitting.

\subsection{Regularity of the Residual Time Series}

Content-driven acquisition tends to be spiky and irregular. If AIM does a good job of estimating content-driven acquisition, then when we remove these estimated values from the aggregate acquisition time series we should be left with a residual time series that is more regular than before. We can even compare multiple tunings of the model, and favor those that yield more regular residuals.

What does regularity mean in this context? In general, acquisition in subscription services follows a predictable seasonal pattern. The more consistent the observed seasonal pattern is across years, the more likely we have successfully removed transient factors. Of course, content launches are not the only transient factors that affect acquisition. Price changes, marketing blitzes, product changes, competitor launches, and world events may all disrupt the seasonal acquisition pattern. For this reason, residual time series will never be completely regular. However, external factors remain constant whether our model has done a good job or a poor job of estimating content-driven acquisition, so all else equal we should expect that better models will yield more regular residuals on average.

\subsection{Reducing Multiple Assignment}

In Section \ref{multi} we detailed an algorithm to minimize multiple assignment at the subscriber level, given a set of aggregate impact estimates. We can go a step further, and reduce multiple assignment more by tuning the model to yield aggregate impact estimates that they collide less at the individual level.

One way to think about multiple assignment is that it is evidence of aggregate overestimation. While it is theoretically possible for two pieces of content to be simultaneously pivotal to a subscriber's decision to join the service, such cases should be rare.  If we see significant multiple assignment despite our efforts to minimize it, it may more likely come from aggregate overestimation of launch impact. We can examine time periods, geographical regions, and pieces of content with high rates of multiple assignment, and look for model tunings that tend to reduce the aggregate estimates associated with them.

\subsection{Other Methods}

For the biggest content launches, there may be a clearly-visible spike in the aggregate number of sign ups. We can use such situations to evaluate how much of the sign up spike is captured by the model, and whether daily estimates are consistent with the aggregate spike. Close alignment between the model and the spike for the biggest pieces of content can build confidence in estimates for smaller pieces of content. 

We can also evaluate how the model behaves when external events impact the aggregate sign up pattern. For example, when COVID-19 hit many streaming platforms saw a surge in acquisition. A well-tuned model will not attribute the rise in signups to whichever content happened to launch at the time, but instead will pass the signup spike into the residual.

Another source of truth that may be available in certain circumstances is content-based acquisition marketing experiments. Suppose a subscription service runs a campaign for a specific piece of content targeted at non-members, and exposure to the campaign is randomized. The randomized campaign will provide us with a causal lift estimate. For a sufficiently powered experiment, we can use AIM to estimate the acquisition impact separately for members of the treatment and control groups of the experiment. This will yield distinct impact estimates for each group. The difference between these group estimates, if the model has performed well, will equal the experimental lift estimate multiplied by the size of the treatment group.

\section{Literature Review}

The majority of prior attribution research focuses on the problem of marketing attribution. A customer purchases a product and the firm wishes to attribute that purchase to some combination of previous events, such as ad impressions. The firm uses cookies or other forms of tracking to observe the timing and nature of previous events to which the customer was exposed. These observations may be incomplete, and may not include the totality of previous events that may have influenced the action. The firm also observes previous events for potential customers who did not purchase. The simplest form of attribution model is rules-based. “Last touch” attribution, a form of rules-based attribution in which the entire purchase is attributed to the most recent touchpoint, gained early popularity not because it was correct, but because it was easy to track on the basis of the referring URL. Berman (2018) shows that last touch attribution incentivizes inefficient oversupply of ad impressions, due to competition among advertisers \cite{berman2018beyond}.

An alternative approach is to observe a corpus of data on touchpoints and conversions, then train a model to determine how much weight to assign each touchpoint. Shao and Li (2011) \cite{shao2011data} develop a multi-touch attribution model of this type. They employ a bagged logistic regression approach in which they train sub-models on subsets of the data, validate misclassification rates against holdout data, then aggregate up into a final model. The bagged approach gives their final model more stable coefficients, which they value for the sake of interpretability by advertisers.  Abhishek et al. (2015) \cite{abhishek2012media} take a different approach, training a Hidden Markov Model (HMM) to simulate the customer’s journey through the conversion funnel, moving from low initial state (“unawareness”) to high final state (“purchase”). The HMM allows them to model how the type and timing of ad exposure alters the transition probabilities between states, which they then use to attribute credit to ads. The authors attempt to use an IV method to account for endogeneity, but ultimately abandon it. Li and Kannan (2014) \cite{li2014attributing} take a related approach by building a hierarchical Bayesian decision model with discrete steps. After estimating the model on hospitality industry data, the authors validate the estimates using a one-week period during which, unlike the training data, paid search was completely turned off. They find that the actual conversion drop (-6.6\%) is smaller than the predicted conversion drop (-7.8\%), suggesting their model did not fully account for substitution across channels. Another data-driven approach is Media Mix Modeling (MMM). MMM tends to use historical data on spend levels and outcomes that is aggregated by time and geography, and runs estimations on these aggregates to tease out relationships. Chan and Perry (2017) \cite{chan2017challenges} write about the challenges of MMM, particularly its inability to account for selection bias and endogenous variation in spend. Chen et al. (2018) \cite{chen2018bias} examine one method to correct for selection bias due to ad targeting in MMM by employing detailed search query data, and show this method leads to improvement in estimates as benchmarked by external experimental data. 

Data-driven models tend to be an improvement over rules-based models, in that their attribution is rooted in empirical evidence. However, lack of exogenous variation means they fall short of capturing the true causal effect of their touchpoints. Ultimately, attribution is a question of causation. The correct value to attribute to an ad is the lift in conversion probability caused by showing the ad, versus the counterfactual of not showing it. For this reason, Randomized Controlled Trials (RCTs) have become a cornerstone of marketing science at Netflix and elsewhere. Sapp et al. \cite{45331} proposed a methodology to simulate a complex causal relationship, to test methods for causal marketing attribution in the absence of RCTs. The Digital Advertising System Simulation (DASS) is a framework for generating simulated customer browsing and ad-viewing histories. DASS is a non-stationary Markov model with three parts: 1) a user path model, 2) an ad serving model, and 3) an ad impact model.  Singh et al. \cite{46901} applied DASS simulation to evaluate five observational estimators: first touch, last touch, linear, matched-pairs data-driven attribution (MP-DDA), and the “upstream data-driven attribution” (MUDDA) method developed in \cite{45766} and extended in \cite{46905}. They find that across a variety of scenarios, MUDDA comes closest to ground truth. MUDDA is an estimator that allows for ad impressions to affect not only conversion probabilities, but also subsequent customer behavior such as search intensity or website visits. Matching on subsequent behavior in a context where such behavior is influenced by the treatment, as is done by the MP-DDA estimator, constitutes a form of post-treatment selection bias which is avoided in MUDDA.

\section{Conclusion}

Understanding acquistion drivers is an extremly important problem for subscription service, as it can  inform content buying decisions, prioritization process in awareness markets and improve demand creating marketing and adaptive decisioning. One driving force that can significantly drive acquistion is the launching of new contents on the service. Yet, the majority of existing literature is limited to marketing attribution research which cannot be directly applied to the subscription domain. In this work, we propose a methodology to better assess the acquisition value of content in a subscription services business. Our model estimates the number of incremental sign ups caused after a content’s launch.  Our methodology combines the intuition behind first-touch attribution (incremental subscribers consume a content quickly), with adjustments for product promotion, subscriber activity and seasonality.  Specifically, our methodology performs cohort analysis to isolate incremental sign ups, using pre-launch new member cohorts as the control group. 

In theory, our methodology can be extended to physical platforms such as  gyms / clubs that also offer membership oriented access for a fee. However we think that the scale and frequency of content launch make our proposed methodology more approapriate to digital platforms.

We introduce a complimentary mixed linear integer programming model to identify specific subscribers who signed up for a content. Our model minimizes the number of subscribers attributed to more than one content, while maximizing the average incremental likelihood of subscribers being attributed across different contents. This attribution enables the subscription services to better understand characteristics of subscribers who join for specific contents, as well as evaluating sign-up dynamics at the content and country grain, and deeper insights about content preferences for new members, including their streaming journey on the subscription service.  

Finally we propose several methods that could be utilized to validate the estimates and calibrate the model. We cannot directly validate our methodology as we never observe how many people join for a specific content.  However we can use external sources of truth to validate our proposed methodology to estimate the incremental subscribers. To validate the estimates, we can measure the correlation between the estimates and external sources of truth such as estimates of the subscriber impact of very large contents, and estimates based on acquisition marketing experiments. If estimates obtained from our methodology are more strongly correlated than currently existing proxies (e.g. first-touch attribution), this indicates our method is an improvement over existing methodologies. We can  also use external sources of truth to calibrate the estimates, either by tweaking the definition of control cohort or other inputs to the model.

\section{Acknowledgments}

The authors wish to thank the many colleagues from Netflix whose contributions enhanced this work, with special thanks to Manping Wang, Steve McBride and Minwoo Choi. We would also like to thank Meghana Bhatt, Natali Ruchansky, Yves Raimond, Ashish Rastogi, and Phil Hebda for their thoughtful review and guidance.
\clearpage
\bibliographystyle{plain}
\bibliography{Refs}
\clearpage
 
\section{Appendices}
 \subsection{Proof of Lemma \ref{lemma_pij}} \label{lemma_pij_proof}

\begin{proof}
	Let's denote $\mathbb{A}_{ij}$ as the event of subscriber  $i$ being incremental to content $j$, $\mathbb{B}_{ij}$ as the event of subscriber  $i$ streaming content $j$ upon signing up, and $\mathbb{A}_{ij}^c$ and $\mathbb{B}_{ij}^c$ as the complement of $\mathbb{A}_{ij}$ and $\mathbb{B}_{ij}$, respectively. Based on Assumption \ref{assumption}, we can write:
	$$P(\mathbb{B}_{ij}|\mathbb{A}_{ij}) = 1,$$
	therefore:
	$$P(\mathbb{A}_{ij})=P(\mathbb{A}_{ij}\cap\mathbb{B}_{ij}).$$
	Using the pre-launch cohort, we can estimate $P(\mathbb{B}_{ij}|\mathbb{A}_{ij}^c)$ using \eqref{eq:ml_model} as $\hat{p}_{ijt}$. We can write $P(\mathbb{B}_{ij}|\mathbb{A}_{ij}^c)$  as 
	$$P(\mathbb{B}_{ij}|\mathbb{A}_{ij}^c)=\frac{P(\mathbb{B}_{ij}\cap\mathbb{A}_{ij}^c)}{P(\mathbb{A}_{ij}^c)}=\frac{P(\mathbb{B}_{ij})-P(\mathbb{B}_{ij}\cap\mathbb{A}_{ij})}{1-P(\mathbb{A}_{ij})}=\frac{P(\mathbb{B}_{ij})-P(\mathbb{A}_{ij})}{1-P(\mathbb{A}_{ij})}.$$ 
	Solving for $P(\mathbb{A}_{ij})$ and replacing $P(\mathbb{B}_{ij}|\mathbb{A}_{ij}^c)$ with $\hat{p}_{ijt}$, we get
	$$P(\mathbb{A}_{ij})=\frac{P(\mathbb{B}_{ij})-\hat{p}_{ijt}}{1-\hat{p}_{ijt}},$$
	which implies
	$$\frac{\partial P(\mathbb{A}_{ij})}{\partial \hat{p}_{ijt}}\le0.$$

\end{proof}
 
\end{document}